\documentclass[prd,nofootinbib,floats,superscriptaddress,eqsecnum,tightenlines,preprintnumbers,11pt]{revtex4}

\usepackage{hyperref}
\usepackage{graphicx}
\usepackage{amsmath,amssymb,amsfonts,amsthm,latexsym}
\usepackage{marginnote}
\usepackage{color}

\def\be{\begin{equation}}
\def\ee{\end{equation}}
\def\beq{\begin{equation}}
\def\eeq{\end{equation}}
\newcommand{\bea}{\begin{eqnarray}}
\newcommand{\eea}{\end{eqnarray}}
\def\bi{\begin{itemize}}
\def\ei{\end{itemize}}
\def\ba{\begin{array}}
\def\ea{\end{array}}
\def\bfig{\begin{figure}}
\def\efig{\end{figure}}

\newcommand{\Ar}{\lambda_1}
\newcommand{\Ard}{\lambda_{1X}}
\newcommand{\Ak}{\lambda_2}
\newcommand{\Aa}{\lambda_3}
\newcommand{\Ag}{\lambda}

\newcommand{\Rf}{ {}^{(4)}\!{R}}
\newcommand{\Rt}{ {}^{(3)}\!{R}}

\newcommand{\Mp}{M_{\rm P}}

\begin{document}

\title{Quadratic DHOST theories revisited}

\author{David Langlois}
\affiliation{Universit\'e de Paris, CNRS, Astroparticule et Cosmologie, F-75006 Paris, France}
\author{Karim Noui}
\affiliation{Institut Denis Poisson, Universit\'e de Tours, Universit\'e d'Orl\'eans, CNRS, UMR 7013, 37200 Tours, France}
\affiliation{Universit\'e de Paris, CNRS, Astroparticule et Cosmologie, F-75006 Paris, France}
\author{Hugo Roussille}
\affiliation{Universit\'e de Paris, CNRS, Astroparticule et Cosmologie, F-75006 Paris, France}\affiliation{Institut Denis Poisson, Universit\'e de Tours, Universit\'e d'Orl\'eans, CNRS, UMR 7013, 37200 Tours, France}

\date{\today}

\begin{abstract}
We present a  novel and remarkably simple formulation of  degenerate higher-order scalar-tensor (DHOST)  theories whose Lagrangian is quadratic in second derivatives of some scalar field.  Using disformal transformations of the metric, we identify a special ``frame'' (or metric)  for which the Lagrangian of quadratic  DHOST  theories  reduces  to  the usual Einstein-Hilbert term plus a few   terms that depend on simple geometric quantities characterizing the  uniform scalar field hypersurfaces. 
In particular, for quadratic DHOST theories in the physically interesting class  Ia, the   Lagrangian simply consists of the Einstein-Hilbert term plus a term proportional to the three-dimensional scalar curvature  of the  uniform scalar field  hypersurfaces. The classification of all quadratic DHOST theories becomes particularly transparent in this geometric reformulation,
which also applies to scalar-tensor theories that are degenerate only in the unitary gauge.
\end{abstract}

\maketitle

\section{Introduction}
Scalar-tensor theories have attracted considerable attention as alternatives to General Relativity.  
As ever more sophisticated models have been considered, special attention was recently devoted to scalar-tensor Lagrangians that contain second order derivatives of the scalar field. A crucial requirement for such theories to be physically relevant is the absence of 
the so-called Ostrogradski ghost \cite{Ostrogradsky:1850fid}, in order to avoid disastrous instabilities (see \cite{Woodard:2015zca,Ganz:2020skf} and references therein). It was believed
for some time that the absence of an Ostrogradski ghost requires the Euler-Lagrange equations to be at most second-order, which explains why the literature was   limited, until a few years ago,  to the study of Horndeski's theories \cite{Horndeski:1974wa}, rediscovered in the guise of the so-called galileons \cite{Nicolis:2008in,Deffayet:2011gz,Kobayashi:2011nu}.
But the discovery of viable theories ``beyond Horndeski'' \cite{Zumalacarregui:2013pma,Gleyzes:2014dya,Gleyzes:2014qga} possessing Euler-Lagrange equations of order higher than two,  
challenged this preconception. 

 It was then realised, in   \cite{Langlois:2015cwa,Langlois:2015skt},  that the absence of the Ostrogradski ghost is automatically guaranteed in Lagrangians whose degeneracy entails constraints that eliminate the potentially  dangerous extra scalar degree of freedom, even if the associated Euler-Lagrange equations are higher-order. This led to the systematic classification of Degenerate Higher-Order Scalar-Tensor (DHOST) theories \cite{Langlois:2015cwa,Langlois:2015skt,Crisostomi:2016tcp,Crisostomi:2016czh,Achour:2016rkg,BenAchour:2016fzp} (see  \cite{Langlois:2018dxi} for a recent review).

Within the family of DHOST theories, one can use the correspondence between different actions that transform into each other  via a disformal transformation of the metric, 
\beq
\label{disformal}
g_{\mu\nu} \longrightarrow \tilde{g}_{\mu\nu} \equiv A(\phi,X) g_{\mu\nu} + B(\phi,X) \phi_\mu \phi_\nu \, ,
\eeq
where $A$ and $B$ are functions of $\phi$, $\phi_\mu \equiv \nabla_\mu \phi$ and $X\equiv \nabla_\mu \phi \nabla^\mu\phi$. 
Provided this transformation is invertible, the actions $\tilde{S}[\phi, \tilde{g}_{\mu\nu}]$ and $S[\phi, g_{\mu\nu}]\equiv\tilde{S}[\phi,  A(\phi,X) g_{\mu\nu} + B(\phi,X) \phi_\mu \phi_\nu]$ are 
physically 
equivalent\footnote{Note that this is only true in absence of matter. Indeed, when matter is included (in the form of a field $\psi_m$), 
the total actions
$S_{tot}[g_{\mu\nu},\phi,\psi_m] = S[g_{\mu\nu},\phi] + S_m[g_{\mu\nu},\psi_m,] $ and 
$\tilde{S}_{tot}[\tilde g_{\mu\nu},\phi,\psi_m] = \tilde{S}[\tilde g_{\mu\nu},\phi] + S_m[\tilde g_{\mu\nu},\psi_m] $
correspond to different physical theories, since matter is minimally coupled to the metric $g_{\mu\nu}$ in the former case
and to $\tilde g_{\mu\nu}$ in the latter case.}. 
One is thus allowed to pick up one of these disformally related metrics $\tilde g_{\mu\nu}$, or ``frame", to write the corresponding theory in a convenient way.

In this article, we identify a ``frame'' where the action of   quadratic DHOST  theories  takes a remarkably simple form, with a natural geometric interpretation based on
the three-dimensional hypersurface $\Sigma_\phi$ where the
scalar field $\phi$ is constant.  Remarkably, we find that all quadratic DHOST theories in class Ia (see \cite{Achour:2016rkg} for the classification), the most interesting for physical applications,
can always be mapped, via a disformal transformation, to the very simple form\footnote{Notice that the action  \eqref{simpleform} is a special case of the so-called spatially covariant gravity actions 
introduced in \cite{Gao:2014soa} and studied
further in \cite{Gao:2014fra}.}
\bea
\label{simpleform1}
S[g_{\mu\nu},\phi] = \int d^4x \sqrt{-g} \left( \frac{1}{2}\Mp^2 \, \Rf + \Ar \, \Rt  \, \right)\,,
\eea
where $\Mp$ is the 4-dimensional Planck mass, $\Rf$  the four-dimensional Ricci scalar, $\Rt$  the three-dimensional Ricci scalar of the hypersurfaces $\Sigma_\phi$
and $\Ar$ an arbitrary function of $X$ and $\phi$. 

More generally, all  quadratic DHOST theories can be written in the form 
\bea
\label{simpleform}
S[g_{\mu\nu},\phi] = \int d^4x \sqrt{-g} \left( \frac{1}{2}\Mp^2 \, \Rf + \Ar \, \Rt + \Ak \, \varepsilon \, K^2 + \Aa \, a^2 \, \right),
\eea
where  $K$ is the trace of the extrinsic curvature  of $\Sigma_\phi$ and $a^2 \equiv a_\mu a^\mu$ the square of  the ``acceleration'' vector of the unit vector normal to $\Sigma_\phi$,  $\varepsilon$
 is the sign\footnote{The sign $\varepsilon$ can be absorbed into a redefinition of
$\Ak$ but we leave it here for later convenience.} 
of $X$, 
while $\Ag_A$ are arbitrary functions of $\phi$ and $X$. Interestingly, quadratic U-degenerate theories introduced   in \cite{DeFelice:2018mkq}, which are degenerate only in the unitary gauge, where the scalar field is a function of time only,  can also be reformulated in the form (\ref{simpleform}).

The formulations  \eqref{simpleform1} and  \eqref{simpleform} can be seen as an analog of  the ``Einstein frame'' for traditional scalar-tensor theories, where the dynamical structure of the gravitational theory is governed by the usual Einstein-Hilbert action with a constant Planck mass. The possibility to reformulate degenerate theories, via disformal transformations, in this ``geometric" frame is the main result of this paper. We also show how the classification of quadratic DHOST theories becomes transparent in this new formulation, as well as the reason behind the instability of several classes of DHOST theories.

 Note that  three-dimensional quantities based on the uniform scalar field hypersurfaces have 
 already been used   in several earlier works, for instance in the context of the EFT of inflation \cite{Cheung:2007st}, of Horava's gravity and its extensions \cite{Blas:2010hb}, in the study of  the cosmology of Horndeski and Beyond Horndeski theories \cite{Gleyzes:2013ooa,Gleyzes:2014rba}. The novelty here is that we combine this three-dimensional formalism with the systematic exploitation of  disformal transformations,
 in order to simplify the description of DHOST theories.

The paper is organized as follows. 
In the next section, we recall the definition of quadratic DHOST, 
define the notion of  ``weakly degenerate'' theories which are degenerate in the unitary gauge (when $\varepsilon=-1$), 
as well as the parametrisations that have been introduced in previous works to describe them.
In Section \ref{section_geometric}, we present the new  formulation of quadratic
DHOST  and weakly degenerate actions. In section \ref{section_DHOST}, we revisit the classification of DHOST theories in this new perspective. We also  show how the action \eqref{simpleform} is related to the very familiar Horndeski action by computing explicitly the disformal transformation which links them. Finally, we present a very simple argument based on this new formulation which proves that
DHOST theories in the classes  II and III are plagued by instabilities (or do not propagate gravitational waves). 
We conclude in section \ref{Sec:conc} with some perspectives.
And we also provide some details and additional results in an Appendix.

\section{Higher-order scalar-tensor theories and degeneracy}

We consider scalar-tensor theories with a quadratic dependence on the second derivatives of the scalar field. Quite generally, their Lagrangian can be written in the form 
\bea
\label{covaction}
L = f(\phi,X) \, \Rf  + \sum_{A=1}^5 \alpha_A(\phi,X) \, L_A \, ,
 \eea
where the elementary quadratic Lagrangians $L_A$ are defined by
\bea
\label{quadLag}
L_1= \phi_{\mu\nu} \phi^{\mu\nu} , \quad
L_2 =\Box \phi^2 , \quad L_3 = \phi^\mu \phi_{\mu\nu} \phi^\nu \Box \phi , \quad
L_4 =( \phi_{\mu\nu} \phi^\nu)^2 \, , \quad
L_5 = (\phi^\mu \phi_{\mu\nu} \phi^\nu)^2\,,
\eea
where $\phi_{\mu\nu} \equiv \nabla_\mu \nabla_\nu\phi$. Note that one can add in the Lagrangian  terms of lower order in $\phi_{\mu\nu}$, which can be written in the form $P(\phi,X)$ (order 0) and $Q(\phi, X) \Box\phi$ (order 1), but we will ignore them here as they do not affect the degeneracy of the action. 

If the six functions $f$ and $\alpha_A$ are arbitrary, one finds generically an extra scalar degree of freedom which leads in general  to an Ostrogradski instability. To avoid the presence of any extra degree of freedom, one needs to impose degeneracy conditions, which define the DHOST theories. For quadratic theories,  the degeneracy imposes three conditions,  derived in \cite{Langlois:2015cwa}, 
\beq
\label{degeneracy}
D_0(X)=0\,, \qquad D_1(X)=0\,, \qquad D_2(X)=0\,, 
\eeq
where the functions $D_0$, $D_1$ and $D_2$ are given  in Appendix \ref{app:degeneracy}.

Instead of working directly with DHOST theories, it will be convenient at this stage to consider a larger family of theories, which are degenerate in a weaker sense as we now explain. Assuming that 
 $\varepsilon={\rm sgn}(X)=-1$, one can introduce the so-called unitary gauge, where the scalar field is spatially uniform and thus depends only on time.
 We will name {\it weakly degenerate}   all  the theories that are degenerate in the unitary gauge. This includes obviously the DHOST theories, which are degenerate in any gauge and thus {\it a fortiori} in the unitary gauge, but also  theories that are degenerate only in the unitary gauge, dubbed U-degenerate theories in \cite{DeFelice:2018mkq}.  Since U-degenerate theories are not DHOST theories, they contain an extra degree of freedom, but this scalar mode is   not propagating in the unitary gauge, as it satisfies an elliptic partial differential equation. 
 
Weakly degenerate theories have been classified in \cite{DeFelice:2018mkq}. For quadratic theories, they satisfy a {\it single} degeneracy condition  (to be contrasted with the {\it three} conditions (\ref{degeneracy}) obeyed by the quadratic DHOST theories), which reads \cite{Langlois:2015cwa}
\begin{equation}
\label{Unitarydegcond}
D_0(X) - X D_1(X) + X^2 D_2(X) = 0 \,.
\end{equation}
Hence only  five out of the six functions in \eqref{covaction} are independent. In \cite{DeFelice:2018mkq}, the unitary degeneracy condition 
\eqref{Unitarydegcond} was solved by expressing the five
functions $\alpha_A$ in terms of $f$ and four independent functions  $\kappa_1$, $\kappa_2$, $\alpha$ and $\sigma$ as follows
\bea
\label{Udeg}
&&\alpha_1 = \kappa_1 + \frac{f}{X} \, , \quad \alpha_2 = \kappa_2 - \frac{f}{X} \, , \quad
\alpha_3 =  \frac{2f - 4X f_X}{X^2}  + 2\sigma \kappa_1 + 2 \left( 3 \sigma - \frac{1}{X}\right) \kappa_2 \, , \\
&&\alpha_4 = \alpha +  2\frac{X(f_X - \kappa_1) - f}{X^2} \, , \quad
\alpha_5 =  \frac{2 f_X - X \alpha }{X^2} + \kappa_1 \left( \frac{1}{X^2} + 3 \sigma^2 - \frac{2\sigma}{X}\right) + \kappa_2 
\left( 3\sigma - \frac{1}{X}\right)^2 . \nonumber
\eea
Quadratic DHOST theories  can be inferred  from weakly degenerate theories by imposing two more additional conditions 
(say $D_0=0$ and $D_1=0$)
and therefore can be parametrised 
in terms of three independent functions only. Quadratic DHOST theories contain several subclasses and the explicit form of their parametrisation depends on the specific subclass considered \cite{Langlois:2015cwa}.

\section{Geometric formulation}
\label{section_geometric}

\subsection{Geometric Lagrangian}
As emphasised in the introduction, the scalar field $\phi$ naturally  induces a preferred slicing of spacetime, which we are going to exploit. Let us 
first introduce various geometric tensors associated with this slicing. 
The intrinsic geometry of any constant $\phi$ hypersurface $\Sigma_\phi$   is characterized by the three-dimensional induced metric 
\bea
\label{3Dgeom}
h_{\mu\nu} \equiv g_{\mu\nu} - \varepsilon \, n_\mu n_\nu \, , \qquad n_\mu \equiv \frac{\phi_\mu}{\sqrt{\vert X \vert }} \, , 
\eea
where $n_\mu$ is the unit vector orthogonal to $\Sigma_\phi$.
One can also introduce the Riemann tensor $\Rt_{\mu\nu\rho\sigma}$ associated with $h_{\mu\nu}$, 
the extrinsic  curvature tensor and the ``acceleration" vector, the components of the latter being respectively given by
\bea
\label{Kanda}
K_{\mu\nu} \equiv  h_\mu^\alpha \, h_\nu^\beta \, \nabla_\alpha n_\beta \, , \qquad
a_\mu \equiv n^\nu \nabla_\nu n_\mu \, .
\eea
By expressing these quantities explicitly in terms of the scalar field and of its derivatives, one notes that both $K_{\mu\nu} $ and $a^\mu$ are linear in  $\phi_{\mu\nu}$ as shown in \eqref{aKphi}. This suggests to rewrite the quadratic Lagrangian in terms of the square of these quantities. 
In the following, we will therefore  examine theories whose action reads
\bea
\label{Lphi}
S[g_{\mu\nu},\phi] = \int d^4x \sqrt{-g} \left( \frac{\Mp^2}{2} \, \Rf + L_\phi \, \right) \,,
\eea
with a Lagrangian term $L_\phi$  of the form
\begin{equation}
\label{fourlambda}
	L_\phi = \Ar \, \Rt + \Ak \, \varepsilon\, K^2 + \Aa \, a^2 + \lambda_4 \, \varepsilon \, K_{\mu\nu} K^{\mu\nu} \, , 
\end{equation}
where $\lambda_A$ are arbitrary functions of $\phi$ and $X$. We have also included a dependence on the three-dimensional scalar curvature  of $\Sigma_\phi$.

It turns out that using only the first three terms, i.e. choosing $\lambda_4=0$, will be sufficient for our purpose as we will explain later. The total action is thus
\bea
\label{simpleaction2}
S[g_{\mu\nu},\phi] = \int d^4x \sqrt{-g} \left( \frac{1}{2}\Mp^2 \, \Rf + \Ar \, \Rt + \Ak \, \varepsilon \, K^2 + \Aa \, a^2 \, \right) \, ,
\eea 
which, using the  relations  in  Appendix \ref{Sec:3surf}, yields  a quadratic action of the form (\ref{covaction}), plus lower order terms also given  in Appendix \ref{Sec:3surf} which we do not discuss here, with the coefficients
\bea
&&f = \frac{\Mp^2}{2} + \Ar \, , \quad
\alpha_1 = \frac{\Ar}{X} \, , \quad
\alpha_2 = \frac{\Ak - \Ar}{X} \, , \nonumber \\
&& \alpha_3 =  \frac{2\Ar - 4X \lambda_{1X} - 2 \Ak}{X^2} \, , \quad
\alpha_4 =  \frac{4X \lambda_{1X} - 2 \Ar  + \Aa}{X^2} \, , \quad
\alpha_5 = \frac{\Ak - \Aa}{X^3} \,.\label{alphalambda}
\eea
It is immediate to check that this quadratic Lagangian is indeed weakly degenerate, as it is of  the form  \eqref{Udeg} with
\begin{equation}
f = \frac{\Mp^2}{2} + \lambda_1 \, , \quad
\kappa_1 = -\frac{\Mp^2}{2X} \, , \quad
\kappa_2 = \frac{\Mp^2 + 2  \Ak}{2X} \, , \quad
\sigma = 0  \, , \quad
\alpha = \frac{\Aa + 2 X {\lambda_{1X}}}{X^2} \,.
\end{equation}

\subsection{Disformal transformations}
We now perform a disformal transformation (\ref{disformal}) of the action (\ref{simpleaction2}), using the formulas  derived in  \cite{Achour:2016rkg} and recalled in  Appendix \ref{app:disformal}. 
The quadratic part of the new action is  still weakly degenerate, i.e. of the form \eqref{Udeg}  with the coefficients
\begin{equation}
\label{disformalUDeg}
	\begin{aligned}
		\sigma &= \frac{A_X}{A}, \quad \kappa_1 = -\frac{\Mp^2}{2X} \frac{A^{3/2}}{\sqrt{A+BX}}, \quad \kappa_2 = \frac{\Mp^2 + 2 \Ak}{2X}\frac{A^{3/2}}{\sqrt{A+BX}}, \\
		\quad f &= \frac{ \Mp^2 + 2 \Ar }{2}  \sqrt{A(A+BX)} , \quad \alpha = {\cal A}(\Ar, \Aa, A, B) \,,
	\end{aligned}
\end{equation}
where we do not write explicity $\cal A$ as it is rather cumbersome (but it can be deduced from the expression (\ref{Expresslambda3}) given below\footnote{Indeed, the expression of $\cal A$ can be obtained directly from \eqref{Expresslambda3} which gives
$$
\alpha = \frac{M^2 A^2(3 \kappa_1 (X \sigma - 1) - 2 X \kappa_{1X})^2}{2 X^3 \kappa_1^3} \Aa  
+ \frac{2 f \sigma (2 + X \sigma) - 2 f_X (1 + 4 X \sigma)}{X} \, .
$$
Then, one substitutes  $f$, $\kappa_1$ and $\sigma$ by their expressions \eqref{disformalUDeg} in terms of $A$, $B$ and $\Ar$ to get $\cal A$ explicitly.}). Interestingly, the new theory 
(\ref{disformalUDeg})
is parametrised by five independent functions, as many functions as required to span the whole family of weakly degenerate theories. 

Conversely, given a weakly degenerate theory defined by the set of functions $(f, \kappa_1, \kappa_2, \sigma, \alpha)$, one can successively
invert the relations \eqref{disformalUDeg} and determine $A$ from $\sigma$, $B$ from $\kappa_1$, $\Ak$ from $\kappa_2$ and $\Ar$ from $f$. The last 
relation yields  $\Aa$, which reads
\bea
\label{Expresslambda3}
\Aa =  \frac{2 (X \alpha - 2 f \sigma (2 + X \sigma) + 2 f_X (1 + 4 X \sigma))X^2 \kappa_1^3}{M^2 A^2(3 \kappa_1 (X \sigma - 1) - 2 X \kappa_{1X})^2} \, .
\eea
This proves that any weakly  degenerate action can be obtained from a disformal transformation of the action \eqref{simpleaction2}. As a conclusion, \eqref{simpleaction2}
provides us with a complete parametrization, up to disformal transformations, of quadratic weakly degenerate theories.

Since the family of weakly degenerate theories is parametrised by five free functions and that disformal transformations depend on two free functions, only three functions are needed to parametrise the geometric frame\footnote{This is the reason why we chose $\lambda_4=0$ in \eqref{fourlambda}. Note that  one could
have made different choices to reduce the number of free functions, e.g. one could impose $\Ak=0$, $\Ar=0$ or another linear relation between the four functions $\lambda_A$. However, one would lose the remarkable simplicity of the geometric frame, specially in class I \eqref{classI}.}. 

Finally, let us note that, although our intuitive reasoning was based on the unitary gauge,  which implicitly assumes that the hypersurfaces $\Sigma_\phi$ are spacelike, i.e. $\varepsilon=-1$, 
the relation between the actions \eqref{covaction} with coefficients 
\eqref{Udeg} and  \eqref{simpleaction2} can be obviously extended to the case $\varepsilon=+1$ even if they can  no longer be 
interpreted as weakly degenerate actions.

\section{Geometric formulation of DHOST theories}
\label{section_DHOST}
All the subclasses of DHOST theories are stable under disformal transformations, as shown in \cite{Achour:2016rkg}. This implies that once  the DHOST theories in the geometric frame have been classified, the classification can immediately be extended to the whole family of DHOST theories, as they are all  generated by disformal transformations from the geometric frame actions. 

\subsection{DHOST theories in class I}
Class I is characterized by the relation $\alpha_1 = - \alpha_2$, which  is equivalent to $\Ak=0$ according to (\ref{alphalambda}). Under this assumption, the first degeneracy condition $D_0=0$  is automatically satisfied and the two other conditions, remarkably, both reduce to $\lambda_3=0$ (see Appendix \ref{app:degeneracy}).
DHOST theories in class I, when expressed in the geometric frame, are thus of the very simple form
\bea
\label{classI}
S_{\rm I} = \int d^4x \sqrt{-g} \left(\frac{1}{2} \Mp^2 \, {\Rf} +  \Ar \, \Rt  \right) \qquad {\rm (class\  I)} \, .
\eea
Within the class I, one finds the subclass Ib characterized by $\alpha_1=f/X$. It is then clear from the first relation in (\ref{alphalambda}) that this subclass corresponds to the action (\ref{classI}) with $\Mp=0$, i.e. a pure three-dimensional curvature term, which does not contain tensor modes \cite{Achour:2016rkg} .

\medskip

Let us discuss further the subclass Ia, which is the most interesting from a physical point of view.  The simplest theory in the geometric frame is obviously General Relativity, with $\Ar=0$. Via disformal transformations, it generates a family of DHOST theories parametrised by two functions, $A$ and $B$. Let us stress that the speed $c_g$ of gravitational waves is modified via a disformal transformation as the causal structure is modified. From an initial value $c_g^2=1+2\lambda_1/\Mp^2$ in the geometric frame\footnote{The quantity $c_g^2$ is easily obtained  from the coefficients of the time derivatives and spatial gradients of the tensor modes in the action. In practice, 
using the Gauss-Codazzi identity \eqref{Gauss-Codazzi}, we express $\Rf$  in terms of $\Rt$ and $ K_{\mu\nu}$ so that the action \eqref{classI} becomes
\bea
S_{\rm I} = \frac{1}{2} \Mp^2  \int d^4x \sqrt{-g} \left[ \left(1 + 2\Ar/\Mp^2 \right)\, {\Rt} + \varepsilon \left( K^2 - K_{\mu\nu} K^{\mu\nu}\right) \right] \, ,
\eea
and then we obtain $c_g^2$  as the ratio of the coefficient of $\Rt$ with the coefficient of $-\varepsilon K_{\mu\nu}K^{\mu\nu}$.}, one gets after disformal transformation
\beq
c_g^2=\left(1+\frac{B X}{A}\right)\left(1+\frac{2\lambda_1}{\Mp^2}\right)\,.
\eeq
As expected, only conformal transformations, i.e. with $B=0$, leave $c_g$ invariant. Consequently,  theories Ia  such that $c_g=c$ (for any solution) can be obtained from a theory with arbitrary $\Ar$ in the geometric frame (for which $c_g\neq c$) via a disformal transformation that compensates the initial detuning of $c_g$ from $c$ so that the final DHOST theory verifies $c_g=c$ (which is equivalent to the condition $\alpha_1=0$ in the original DHOST formulation \cite{Langlois:2017dyl,Crisostomi:2017lbg,deRham:2016wji}). 

Note that the DHOST theories that can play the role of dark energy while satisfying both  $c_g=c$ and the GW decay constraint \cite{Creminelli:2018xsv}, as suggested by the GW170817 observation \cite{TheLIGOScientific:2017qsa}, correspond to the theories generated via  conformal transformations ($B=0$) from General Relativity in the geometric frame, i.e. with $\lambda_1=0$. The cosmology of such theories has been explored in  \cite{Frusciante:2018tvu}. However, since  the LIGO-Virgo measurements probe wavelengths many orders of magnitude smaller than cosmological scales, these constraints do not necessarily apply on cosmological scales (see e.g. \cite{deRham:2018red}), leaving the other  DHOST theories still  relevant for cosmology  \cite{Boumaza:2020klg}.

In addition to the geometric  frame introduced in the present work, another convenient frame for the theories Ia,  is the ``Horndeski frame'' where the corresponding action is of the Horndeski form (up to lower order terms), i.e. 
\begin{equation}
\label{horny}
	S_H[g_{\mu\nu},\phi] = \int d^4x \sqrt{-g} \left( F \, \Rf + 2 F_X (L_1- L_2)  \right) \,.
\end{equation}
It is thus interesting to derive explicity the disformal transformation that relates these two frames. Given any (quadratic) Horndeski action, characterized by a function $F(\phi,X)$, one can define 
\begin{equation}
\label{lambdaF}
	\Ar = - \frac{\Mp^2}{2} + \frac{2}{\Mp^2}  F(F-2 X  F_X )  \, ,
\end{equation}
and verify  that the disformal transformation of \eqref{classI} with $A = \text{sg}(F-2XF_X) \in \{ +1,-1\}$  and
\bea
A B = \frac{\Mp^4}{4 X (2XF_X-F)^2} - \frac{1}{X}\,,
\eea
gives exactly \eqref{horny}. Note that we have necessarily $2XF_X-F\neq 0$ since we are not in the class Ib.

\subsection{DHOST theories in the classes II and III}
\subsubsection{Classification}
Let us first discuss DHOST theories in the class III, characterized by $f=0$, i.e. 
\beq
 \Ar =-\frac{\Mp^2}{2}\qquad {\rm (class\  III)} \, .
 \eeq
 If $\Mp\neq 0$, the degeneracy conditions $D_1=0$ and $D_2=0$ then imply
 \beq
 \Aa=0\quad {\rm (subclass\ IIIa)}\,\qquad {\rm or}\qquad  \Ak =-\frac{\Mp^2}{3}\quad{\rm (subclass \ IIIb)}\,,
 \eeq
 corresponding  to the subclasses IIIa and IIIb respectively. Finally, the case $\Mp=0$ and thus $\Ar=0$, with $ \Aa$ and $\Ak$ free, corresponds to the subclass IIIc, which does not contain tensor modes.
 
 \medskip
 
 Let us now turn to the class II,  which contains the theories that are neither in class I or in class III. From the degeneracy conditions \eqref{eq:degeneracy-geometry}, one finds that 
 the class IIa is characterized by
  \bea
\label{formAr}
  \Ar + \frac{\Mp^2}{2}=f(X) = \xi \sqrt{ \vert X \vert}  \, ,  \qquad \Aa=0\,,\qquad {\rm (subclass\  IIa)} \, .
\eea
where $\xi$ is a constant. The class IIb is characterized by two free functions, $\Ar$ and $\Ak$, and the  conditions
 \beq
 \Mp=0\,, \qquad \Aa=2\frac{(\Ar-2X {\Ar}_X)^2}{\Ar} \qquad {\rm (subclass\  IIb)} \, .
 \eeq
 which solve the degeneracy conditions.  
 
 As a consequence, any DHOST theory  in the class II can be described by the action
\bea
S_{II}[g_{\mu\nu},\phi] & =&  \xi   \int d^4x \sqrt{-g\vert X \vert}  \, \Rt - \frac{\Mp^2}{2} \varepsilon \int d^4x \sqrt{-g} \left( K_{\mu\nu} K^{\mu\nu} -\left(1 -  {\mu}_2 \right) K^2\right) \\
& =&  \xi   \int d^4x \sqrt{\vert h \vert}  \, \Rt - \frac{\Mp^2}{2} \varepsilon \int d^4x \sqrt{-g} \left( K_{\mu\nu} K^{\mu\nu} -\left(1 -  \mu_2 \right) K^2\right) \, , \label{class2lambda}
\eea
where we used the relation \eqref{det} between the determinants $g$ and $h$ and we introduced the dimensionless parameter
$\mu_2 \equiv {2\lambda_2}/{\Mp^2}$.
Interestingly the three-dimensional Ricci term in this action  reduces exactly to the three-dimensional Einstein-Hilbert Lagrangian
for the metric $h_{\mu\nu}$ but integrated over the four-dimensional space-time.

\subsubsection{Instabilities}

The geometric frame reformulation \eqref{class2lambda}  of  DHOST theories in class II is particularly convenient to see that these theories are plagued by instabilities, as originally shown in \cite{Langlois:2017mxy}. Indeed, if one considers solutions  with $\varepsilon=-1$  in  the unitary gauge,  spatial gradients  in the equations of motion of the (scalar and tensor) fields can only originate from  the $\Rt$ term in the action. 

 Let us consider a homogeneous and isotropic background, with scale factor $a(t)$
and scalar field  $\phi(t)$. In the unitary gauge, the perturbations about such a background are fully encoded in the scalar perturbation
$\zeta$ and the tensor perturbation $\gamma_{ij}$ of the three-dimensional metric $h_{ij}$,  
\bea
\label{cosmodec}
h_{ij} = a^2(t) e^{2 \zeta}  ( \delta_{ij} + \gamma_{ij}) \, , 
\eea
where latin letters $(i,j,\cdots)$ hold for spatial indices. When we substitute this expression  into the  the $\Rt$ term  which appears in  
\eqref{class2lambda}, we obtain at quadratic order in the perturbations
\bea
\label{quadraticR3}
\int d^4x \sqrt{h}  \, \Rt  =  \int d^4x \, a \,  \left( -\frac{1}{4} \partial_k \gamma_{ij} \,  \partial^k \gamma^{ij} +   \partial_i \zeta \,  \partial^i \zeta  + {o} (\gamma^2, \zeta^2, \gamma \zeta)\right) \, .
\eea
Hence, we immediately see that the gradient term of the scalar mode $\zeta$ has an opposite sign compared to the tensor modes and therefore there will be necessarily a gradient
instability either in the tensor sector or in the scalar sector\footnote{In the case where $\varepsilon=+1$, one cannot take the unitary gauge anymore, but instead one could fix $\phi$ to be one of the three spatial coordinates. Hence, a similar analysis would lead to a quadratic action of the form \eqref{quadraticR3} with $(i,j,k)$ three-dimensional space-time indices while the terms with $K_{\mu\nu}$ in \eqref{class2lambda} would involve now (space-like) gradients of the fields only. Therefore, the only kinetic terms of the scalar field and of  the tensor modes would be contained in  \eqref{quadraticR3} and we clearly see that they have opposite signs, which means that there would be a ghost instability.}. Thus, we recover very easily  the result that DHOST theories in class II are unstable \cite{Langlois:2017mxy}. We see that this instability
is closely related  to the form \eqref{formAr} of $\Ar$ which is necessary to select a DHOST theory among weakly degenerate theories.

\section{Conclusion}
\label{Sec:conc}
In this work, we have  obtained a strikingly simple reformulation of quadratic DHOST theories (and of weakly degenerate theories), based on a Lagrangian involving a few geometrical terms associated with the three-dimensional constant $\phi$ hypersurfaces, in addition to the standard Einstein-Hilbert term. This geometric frame action describes only a subset of DHOST theories but the rest of the family can be ``generated'' from this subset via disformal transformations. 

Moreover, since the various subclasses of DHOST theories are stable under disformal transformations, it is sufficient to classify the subset of theories in the geometric frame to automatically  generalise this classification to the whole family. A compelling illustration is given by the subclass Ia, the most interesting class from a phenomenological perspective, which includes Horndeski's theories. In the original classification of DHOST theories, this subclass is parametrised by the three functions $f$, $\alpha_1$ and $\alpha_3$, while the other functions are expressed in terms of these, with rather ugly expressions for $\alpha_4$ and $\alpha_5$. By contrast, in this new geometric perspective, the subclass Ia arises from an geometric frame action that depends on a single function $\Ar$  multiplying the three-dimensional curvature, all the other theories being obtained via disformal transformations. The subclass Ia is thus parametrised by the three functions $\Ar$, $A$ and $B$. One can proceed similarly for the other DHOST subclasses, as well as for all the quadratic weakly degenerate theories which are also of the form \eqref{simpleaction2}.

As we have argued in this article, the geometric frame perspective  is appealing to understand  and  analyse the underlying dynamical structure of DHOST theories, similarly to the Einstein frame in traditional scalar-tensor theories. However we should stress that, in the presence of matter, it is much more convenient in general, from a practical point of view, to stick to the physical frame where matter is minimally coupled to the metric rather than to move to the geometric frame. Indeed, matter would be disformally coupled to geometric frame metric, with the functions $A$ and $B$ usually defined only implicitly. This would make the calculations in this frame very cumbersome. For concrete applications, it is thus more appropriate to work directly with the original formulation and to add matter  minimally  coupled to the metric in \eqref{covaction}. Scalar-tensor theories that are disformally related thus correspond to physically distinct theories since matter is minimally coupled to both theories. 

By contrast, the geometric frame approach should  be useful to get a better intuitive understanding  of DHOST theories as well as for their classification, or to study their generic properties invariant under disformal transformations, such as the instabilities in classes II and III. 
In this respect, it would be interesting to extend the geometric frame description to include the cubic DHOST theories, whose classification in the standard formulation is even more involved than for the quadratic case. We plan to explore this question in a future work.

\acknowledgements{We would like to thank Christos Charmousis for very instructive discussions. KN acknowledges support from the CNRS grant 80PRIME and thanks ENS for their hospitality.}

\appendix

\section{Uniform scalar field hypersurfaces $\Sigma_\phi$ and the geometric frame}
\label{Sec:3surf}

This section is devoted to recall  useful geometrical properties of  the uniform scalar field hypersurfaces
$\Sigma_\phi$ and its consequences of the reformulation of DHOST and weakly degenerate theories. 

\subsection{Geometry of the hypersurfaces $\Sigma_\phi$ }
The geometry of $\Sigma_\phi$ is fully characterized by the induced metric
$h_{\mu\nu}$ and its four-dimensional normal $n_\mu$ \eqref{3Dgeom},
\bea
h_{\mu\nu} \equiv g_{\mu\nu} - \varepsilon n_\mu n_\nu \, , \qquad n_\mu \equiv \frac{\phi_\mu}{\sqrt{\vert X \vert}} \, , \qquad
\varepsilon \equiv \text{sg}(X) \, .
\eea
From these two tensors one can construct the extrinsic  curvature tensor and the ``acceleration" vector, their components  being respectively given by
\bea
\label{Kanda}
K_{\mu\nu} \equiv  h_\mu^\alpha \, h_\nu^\beta \, \nabla_\alpha n_\beta \, , \qquad
a_\mu \equiv n^\nu \nabla_\nu n_\mu \, .
\eea
 
  Their explicit expressions in terms of $\phi$, its first derivatives
$\phi_\mu$ and its second derivatives $\phi_{\mu\nu} $ can be easily computed:
\bea
\label{aKphi}
a_\mu =  -\frac{1}{2\vert X \vert} h_{\mu\nu} X^\nu\, ,  \quad
K_{\mu\nu} =   \frac{1}{\sqrt{\vert X \vert}} \left[ 
\phi_{\mu\nu} + \frac{\phi^\alpha X_\alpha}{2X^2} \phi_\mu \phi_\nu - \frac{1}{2X} \left( \phi_{\mu} X_{\nu} + \phi_\nu X_\mu \right) \right] ,
\eea
where $X_\mu \equiv \partial_\mu X = 2 \phi_{\mu\nu}  \phi^\nu$.

\medskip

The induced three-dimensional Riemann tensor is also of great importance. It is given by
\bea
\Rt_{\mu\nu\rho\sigma} = h^\alpha_\mu h^\beta_\nu h^\gamma_\rho h^\delta_\sigma \, \Rf_{\alpha \beta \gamma \delta} + \varepsilon
(K_{\mu\rho} K_{\nu \sigma} - K_{\nu \rho} K_{\mu \sigma}) \, .
\eea
This relation enables us to compute the three-dimensional Ricci tensor and  the Ricci scalar which is given by the Gauss-Codazzi relation,
\bea
\label{Gauss-Codazzi}
\Rt  =  \Rf - \varepsilon \left[K^2 - K_{\mu\nu} K^{\mu\nu} - 2 \nabla_\mu (a^\mu - K n^\mu) \right] \, .
\eea
To obtain a more explicit form, we can use the following equations,
\bea
\varepsilon \, K^2 & = &  \frac{1}{X} L_2 - \frac{2}{X^2} L_3 + \frac{1}{X^3} L_5 \, , \\
\varepsilon \,  K_{\mu\nu} K^{\mu\nu} & = &   \frac{1}{X} L_1 - \frac{2}{X^2} L_4 + \frac{1}{X^3} L_5 \, , \\
a^2 & = & \frac{1}{X^2} L_4 - \frac{1}{X^3} L_5 \, ,
\eea
where $L_A$ are the elementary quadratic Lagrangians \eqref{quadLag}, together with the relation
\bea
\varepsilon \int d^4 x \, \sqrt{-g} \, f \, \nabla_\mu (a^\mu - K n^\mu) = 
\int d^4 x \, \sqrt{-g} \left[ - \frac{2 f_X}{X}  L_3 + \frac{2 f_X}{X}  L_4 + \frac{ f_\phi }{X} ( X\Box \phi - \phi_{\mu\nu} \phi^\mu \phi^\nu)  \right] \, ,
\eea 
where $f$ is an arbitrary function of $\phi$ and $X$, $f_X$ its derivative with respect to $X$ and $f_\phi$ its derivative with respect to $\phi$.

\medskip

Finally, one can express the determinant $g$ of the metric $g_{\mu\nu}$
in terms of the determinant $h$ of $h_{\mu\nu}$ and $X$ as follows
\bea
g & = &\frac{1}{24}\epsilon^{\mu_1\nu_1\rho_1\sigma_1} \epsilon^{\mu_2\nu_2\rho_2\sigma_2} g_{\mu_1\mu_2}
g_{\nu_1\nu_2}g_{\rho_1\rho_2}g_{\sigma_1\sigma_2} \nonumber \\
&=& \frac{1}{6 X} \epsilon^{\mu_1\nu_1\rho_1\sigma_1} \epsilon^{\mu_2\nu_2\rho_2\sigma_2} h_{\mu_1\mu_2}
h_{\nu_2\nu_2}h_{\rho_1\rho_2} \phi_{\sigma_1} \phi_{\sigma_2} = \frac{h}{X} \, ,\label{det}
\eea
which follows from the very definition of the determinant and $\epsilon_{\mu\nu\rho\sigma}$ is the fully antisymmetric four-dimensional
tensor.

\subsection{DHOST theories in the geometric frame}
As a consequence, the simplified action \eqref{simpleform}
\bea
S[g_{\mu\nu},\phi] = \int d^4x \sqrt{-g} \left( \frac{1}{2}\Mp^2 \, \Rf + \Ar \, \Rt + \Ak \, \varepsilon \, K^2 + \Aa \, a^2 \, \right),
\eea
can be reformulated as a sum of a quadratic action in the more usual form \eqref{covaction} with the coefficients,
\bea
&&f = \frac{\Mp^2}{2} + \Ar \, , \quad
\alpha_1 = \frac{\Ar}{X} \, , \quad
\alpha_2 = \frac{\Ak - \Ar}{X} \, , \nonumber \\
&& \alpha_3 =  \frac{2\Ar - 4X \lambda_{1X} - 2 \Ak}{X^2} \, , \quad
\alpha_4 =  \frac{4X \lambda_{1X} - 2 \Ar  + \Aa}{X^2} \, , \quad
\alpha_5 = \frac{\Ak - \Aa}{X^3} \, ,\label{alphalambdaA}
\eea
 supplemented with  a k-essence and a cubic galileon action given by
\bea
\label{bounterm}
 \int d^4x \sqrt{-g} \, \frac{2 \lambda_{1\phi}}{X}(X \Box \phi - \phi_{\mu\nu} \phi^\mu \phi^\nu )= 
  \int d^4x \sqrt{-g} \, \left( (2 X \beta_X + \beta) \Box \phi + \beta_\phi X \right) \ ,
  \label{eq:k-essence-geometry}
\eea
with $X\beta =\lambda_{1\phi}$. The identity \eqref{bounterm} follows from the relation
\bea
2  \int d^4x \sqrt{-g} \, \beta_X \phi_{\mu\nu} \phi^\mu \phi^\nu = -  \int d^4x \sqrt{-g} \, \left( X \beta_\phi + \beta \Box \phi \right) \, ,
\label{eq:rel-Y-boxphi}
\eea
for any function $\beta(\phi,X)$.

\subsection{Geometric formulation of the cubic Galileon}
As we have shown that a quadratic DHOST action can be rewritten in a more geometrical way, it is natural to look for a way to replace the cubic galileon term $Q(\phi, X) \Box\phi$ by a combination of $K_{\mu\nu}$ and $a_\mu$ as well. The only combination of these objects that is linear in $\phi$ is the trace of $K_{\mu\nu}$. Therefore, we now consider the action
\begin{equation}
\label{Spikappa}
	S[g_{\mu\nu}, \phi] = \int d^4x \sqrt{-g} \left(\nu_0 + \nu_1 K \right) \,,
\end{equation}
where $\nu_0$ and $\nu_1$ are two arbitrary functions of $X$ and $\phi$. Using \eqref{aKphi}, this action can be written as
\begin{equation}
\label{defofalphainS}
		S[g_{\mu\nu}, \phi] = \int d^4x \sqrt{-g} \left[\nu_0  + \alpha (\phi^\mu\phi^\nu\phi_{\mu\nu} - X \Box\phi)\right] \,, \quad \alpha \equiv -  \frac{\varepsilon \nu_1}{{\vert X \vert^{3/2}}}  \, .
\end{equation}
Using \eqref{eq:rel-Y-boxphi}, we can show that, for any arbitrary function $\alpha(\phi,X)$,
\begin{equation}
		\int d^4x \sqrt{-g} \, \alpha (\phi^\mu\phi^\nu\phi_{\mu\nu} - X \Box \phi) = -\frac{1}{2} \int d^4x \sqrt{-g} \left[ {X} A_\phi + \left({A} + 2\alpha  X\right) \Box\phi\right] \,,
\end{equation}
where  $A$ is such that $A_X = \alpha$. As a consequence, the action \eqref{Spikappa} can be written as
\begin{equation}
	S[g_{\mu\nu}, \phi] = \frac{1}{2}\int d^4x \sqrt{-g} \left(2 \pi - {X} A_\phi - \left({A} + 2 \alpha X\right) \Box\phi\right) \,.
\end{equation}
One recovers the well-known k-essence and cubic galileon terms associated with the functions $P$ and $Q$ given by,
\begin{equation}
\label{Cubicgeometric}
		\begin{aligned}
			P(\phi, X) = \nu_0 - \frac{X}{2} A_\phi \,, \qquad
			Q(\phi, X) = - \frac{1}{2}\left(A + 2\alpha X\right) \,.
		\end{aligned}
\end{equation}
If the quadratic action \ref{simpleform} is also taken into account, it gives a linear contribution given by \ref{eq:k-essence-geometry}. One must therefore replace $\alpha$ in \eqref{defofalphainS} by 
\begin{equation}
	\alpha =  - \frac{\varepsilon \nu_1}{\vert X\vert^{3/2}}   + \frac{2\lambda_{1\phi}}{X}  \,.
\end{equation}

\section{Equations of motion in the geometric frame}
\label{app:eom}
The expression and geometrical interpretation of DHOST theories are much simpler in this novel geometric frame than in the  Horndeski frame. The equations of motion are  also simpler and can be formulated  in geometrical terms which could help finding solutions. 

In this appendix, we compute the equations of motion of the Class I action \eqref{classI} 
\bea
S = \int d^4x \sqrt{-g} \left(\frac{1}{2} \Mp^2 \, {\Rf} +  \Ar \, \Rt  + \nu_1 \, K \, + \, \nu_0 \right) \,,
\eea
to which we added (for purposes of generality) 
a  k-essence term $S_{\nu_0}$ associated with the function $\nu_0(\phi,X)$ and a  K-term $S_{\nu_1}$ associated
with the function $\nu_1(\phi,X)$ \eqref{Spikappa}. As we proved in the appendix \ref{Sec:3surf}, the K-term can be equivalently reformulated as a cubic Galileon up to k-essence terms. 

Using the formulae
\bea
\delta X & = & 2 \phi^\mu \partial_\mu \delta \phi - \phi^\mu \phi^\nu \delta g_{\mu\nu} \, , \\
\delta h_{\mu\nu} & = & \delta g_{\mu\nu} + \frac{\delta X}{X^2} \phi_\mu \phi_\nu - \frac{\phi_\mu \partial_\nu \delta \phi + \phi_\nu \partial_\mu \delta \phi}{X} \, ,
\eea
for the infinitesimal variations of $X$ and $h_{\mu\nu}$ together with 
the relation $h= X g $ \eqref{det} between the determinants  $h= \text{det} (h_{\mu\nu}) $ and $g= \text{det} (g_{\mu\nu}) $, we show by a direct calculation that the equation of motion  for the metric takes the standard form,
\bea
\frac{\Mp^2}{2} \, {}^{(4)}{G}_{\mu\nu} &=&   T_{\mu\nu}^{(\Ar)} +  T_{\mu\nu}^{(\nu_1)} +  T_{\mu\nu}^{(\nu_0)} 
\eea
where  $T_{\mu\nu}^{(\nu_0)} $ and $T_{\mu\nu}^{(\nu_1)}$  are the usual   stress-energy tensors associated to the actions
$S_{\nu_0}$ and $S_{\nu_1}$ respectively,
\bea
T_{\mu\nu}^{(\nu_0)}  =  \frac{\nu_0}{2} g_{\mu\nu } - \nu_{0X} \phi_\mu \phi_\nu \, , \qquad
T_{\mu\nu}^{(\nu_1)} = \frac{1}{2q}\left( \nu_{1\phi} - 2\nu_{1X} \Box \phi \right)  \phi_\mu \phi_\nu   \, ,
\eea
where $q \equiv {\vert X \vert^{1/2}}$, while $T_{\mu\nu}^{(\Ar)}$ is the stress-tensor energy associated to the three-dimensional Ricci term in \eqref{classI}
\bea
T_{\mu\nu}^{(\Ar)} &=&  -q \left[ \mu_1 \, {}^{(3)} \! G_{\mu\nu} + \mu_{1X}  \Rt \, \phi_\mu \phi_\nu + (h_{\mu\nu} {}^3 \Box -{}^3 \nabla_\mu {}^3 \nabla_\nu) \mu_1 \right]\, , 
\eea
where ${}^3 \nabla_\mu$ is the 3-dimensional
covariant derivative with respect to $h_{\mu\nu}$ and  we introduced the notation $\mu_1 \equiv \lambda_1/q$
for simplicity.

Even though the equation of motion for the scalar field is redundant as it can be deduced from the previous one, 
it is nonetheless useful to  give its expression which takes the form
\bea
\nabla_\mu J^\mu + \Phi = 0 \,,
\eea
where the curent $J^\mu$ and the source $\Phi$ are given by,
\bea
J^{\mu} = J^{\mu}_{(\lambda_1)} + J^{\mu}_{(\nu_1)} + J^{\mu}_{(\nu_0)} \, , \qquad
\Phi = \Phi_{(\lambda_1)} + \Phi_{(\nu_1)} + \Phi_{(\nu_0)} \, ,
\eea
and each components of $J^{\mu}$ and $\Phi$ are given by,
\bea
&&J^{\mu}_{(\lambda_1)}  =    \lambda_{1X}   \Rt  \phi^\mu \, , \quad
J^{\mu}_{(\nu_1)} =  \frac{1}{2q}( 2q \nu_{1X} K - {\nu_{1\phi}})\phi^\mu + q \nu_{1X} a^\mu   \, , \quad
J^{\mu}_{(\nu_0)} =  \nu_{0X} \phi^\mu  \, , \\
&& \Phi_{(\lambda_1)}  = -\frac{\lambda_{1\phi}}{2}  \, \Rt \, , \quad
\Phi_{(\nu_1)} = \frac{1}{2 q} \left[ X \nu_{1\phi\phi}  + 2X \nu_{1\phi X} (\Box \phi - q K)\right]    \, , \quad
\Phi_{(\nu_0)} = -\frac{\nu_{0\phi}}{2}   \, .
\eea
If the theory is shift symmetric, i.e. the functions $\Ar$, $\nu_1$ and $\nu_0$ do not depend on $\phi$, we recover the well-known fact that the 
equation of the scalar field reduces to a conservation equation for the curent $J^\mu$. 

Notice that the equation for the metric involves third order derivatives of the scalar field 
and, similarly,  the equation for the scalar field involves third order derivatives of the metric components. This is expected 
as the action is not formulated in the Horndeski frame. 
However, in the case where $\varepsilon=-1$, these higher order terms are all spatial derivatives which is consistent with the fact that there
is no ghost propagating in the theory.  These equations have a  simple form compared to Horndeski theories. This might help us in finding solutions. 

\medskip

Let us see how the conditions for having stealth solutions (see \cite{Babichev:2013cya} for the first stealth solutions in Horndeski theories) 
are formulated in the geometric frame. 
We assume that $X$ is a constant $X_0$. Thus, the equations 
of motion simplify drastically and we have (for the metric only)
\bea
{\Mp^2} \, {}^{(4)}\!{G}_{\mu\nu}   - \nu_0 \, g_{\mu\nu }  + 2 q \, \mu_1 \, {}^{(3)}\!G_{\mu\nu} +2 \left( q \, \mu_{1X} \, \Rt + \nu_{0X}+ \frac{\nu_{1X}}{q} \Box \phi \right) \phi_\mu \phi_\nu   = 0 \, .
\eea
Now, we see that, when  the theory satisfies the following conditions\footnote{Notice that these conditions are consistent with those obtained in \cite{Takahashi:2020hso} for DHOST theories when their actions are
	written in the usual form \eqref{covaction} (supplemented with a k-essence term and a cubic galileon term), 
	\bea
	P + 2\Lambda f = 0 \, , \quad
	P_X + \Lambda (4 f_X - X \alpha_{1X}) = 0 \, , \quad Q_X = 0 \, , \quad
	\alpha_1=0 \, , \quad
	\alpha_3 + 2 \alpha_{1X} = 0 \, .
	\eea
	When we replace the coefficients $\alpha_A$ by their expressions \eqref{alphalambda} in terms of $\lambda_1$ and also, $P$
	and $Q$ by their expressions in terms of $\nu_0$ and $\nu_1$ \eqref{Cubicgeometric}, we recover immediately
	the stealth conditions \eqref{stealthcond}. In particular, we show that 
	$Q_X = \nu_{1X}/q$, 
	which immediately implies the equivalence between the conditions $Q_X=0$ and $\nu_{1X}=0$.},
\bea
\label{stealthcond}
\mu_1(X_0) = 0 \, , \quad \mu_{1X}(X_0) = 0 \, , \quad \nu_{0X}(X_0) = 0 \,, \quad \nu_{1X}(X_0)=0 \, ,
\eea
it admits stealth solutions which satisfy the usual Einstein equation for general relativity, 
\bea
{}^{(4)}\! {G}_{\mu\nu}   + \Lambda g_{\mu\nu } =0 \, , \qquad
\Mp^2 \Lambda + {\nu_0(X_0)} = 0  \, .
\eea
The conditions \eqref{stealthcond} are sufficient but not necessary for the existence of stealth solutions. The theory could admit stealth Schwarzschild solutions for instance without satisfying these conditions as shown in  \cite{Takahashi:2020hso} (and references therein) for instance .

\medskip

\section{Degeneracy conditions for quadratic DHOST theories}
\label{app:degeneracy}

In this section, we recall the degeneracy conditions for quadratic DHOST theories derived in \cite{Langlois:2015cwa}, and then give their expression in terms of the parameters $\Ar$, $\Ak$ and $\Aa$.
The three degeneracy conditions for the quadratic DHOST theories are
\beq
D_0(X)=0\,, \qquad D_1(X)=0\,, \qquad D_2(X)=0\,, 
\eeq
with
\begin{equation}
	\begin{aligned}
		D_0(X) &= -4(\alpha_1 + \alpha_2) \left[X f(2\alpha_1 + X \alpha_4 + 4 f_X) - 2f ^2 - 8 X^2 f_X^2\right] \,,\\[1ex]
		D_1(X) &= 
		\begin{aligned}[t]
			&4 \left[X^2 \alpha_1(\alpha_1 +3 \alpha_2) - 2f^2 - 4 X f \alpha_2\right] \alpha_4+ 4 X^2 f(\alpha_1 + \alpha_2)\alpha_5 \\
			&+ 8 X \alpha_1^3 - 4 (f + 4Xf_X - 6X\alpha_2) \alpha_1^2 - 16(f + 5Xf_X) \alpha_1 \alpha_2 \\
			&
			+ 4X(3f - 4X f_X) \alpha_1 \alpha_3 - X^2 f \alpha_3^2 + 32 f_X(f + 2X f_X) \alpha_2
			\\
			&
			-16 f f_X \alpha_1 - 8 f (f-Xf_X)\alpha_3 + 48 f f_X^2 \,,
		\end{aligned}
		\\[1ex]
		D_2(X) &= 
		\begin{aligned}[t]
			&4 \left[2f^2 + 4Xf\alpha_2 - X^2\alpha_1(\alpha_1+3\alpha_2)\right] \alpha_5 + 4 \alpha_1^3 + 4(2\alpha_2 - X \alpha_3 - 4 f_X) \alpha_1^2 \\
			&
			+ 3X^2 \alpha_1 \alpha_3^2 
			-4Xf\alpha_3^2 + 8(f+Xf_X)\alpha_1\alpha_3 - 32 f_X \alpha_1 \alpha_2 \\
			&
			+ 16 f_X^2 \alpha_1 + 32 f_X^2 \alpha_2 - 16 f f_X \alpha_3 \,.
		\end{aligned}
	\end{aligned}
	\label{eq:degeneracy-DHOST}
\end{equation}

These expressions simplify quite a lot when they are written in term of $\Ar$, $\Ak$ and $\Aa$, substituting  \eqref{alphalambda}: 
\begin{equation}
	\begin{aligned}
		D_0(X) &= \frac{2 \Ak}{X} \left[ \left({\Mp^2} + 2 \Ar\right) \left({\Mp^2} + 2 \Ar - \Aa-8X  \Ard \right)+ 16X^2 \Ard^2 \right] \,,\\[1ex]
		D_1(X) &= - \frac{2 \Mp^4}{X^2} \Aa + \frac{2 \Ak}{X^2} \left[8 \Ar^2 - 5 \Mp^2 \Aa + 4 \Ar(2 \Mp^2 - \Aa - 8 X \Ard) + 2(\Mp^2 - 4X\Ard)^2\right] \,,\\
		D_2(X) &=  - \frac{2 \Mp^4}{X^2} \Aa + \frac{2 \Ak}{X^2} \left[4 \Ar^2 - 4 \Mp^2 \Aa + 2 \Ar(2 \Mp^2 - \Aa - 8 X \Ard) + (\Mp^2 - 4X\Ard)^2\right] \,.\\
	\end{aligned}
	\label{eq:degeneracy-geometry}
\end{equation}

Weakly degenerate theories satisfy the single condition \cite{Langlois:2015cwa} 
\begin{equation}
	D_0(X) - X D_1(X) + X^2 D_2(X) = 0 \,.
\end{equation}

\section{Formulae for the disformal transformations}
\label{app:disformal}

In this appendix, we recall how a quadratic higher-order scalar-tensor action transforms under disformal transformations. 
Notations are based on those introduced in the paper \cite{Achour:2016rkg}. Notice that the importance of disformal transformations in higher-order scalar-tensor theories 
has been realized first in \cite{Zumalacarregui:2013pma}.

We consider the quadratic higher-order scalar-tensor action
\begin{equation}
    \tilde{S}[\tilde{g}_{\mu\nu},\phi] = \int d^4x \sqrt{-\tilde{g}} \left( \tilde{f}(\tilde{X}) \, \Rf  + \sum_{A=1}^5 \tilde{\alpha}_A(\tilde{X}) \, L_A \right) \,,
\end{equation}
where $L_A$ are the elementary quadratic Lagrangians \eqref{covaction} and  $\tilde{X} \equiv \tilde{g}^{\mu\nu} \phi_\mu \phi_\nu$. For simplicity, we assume that the theory is shift-symmetric. 

Disformal transformations of the metric  \ref{disformal} induce a disformal transformation of the action according to
\bea
 \tilde{S}[\tilde{g}_{\mu\nu},\phi]  \; \longrightarrow \; S[g_{\mu\nu},\phi] = \tilde{S}[A(X) g_{\mu\nu} + B(X) \phi_\mu \phi_\nu] \, ,
\eea
where the quadratic part of $S[g_{\mu\nu},\phi] $ is of the form,
\begin{equation}
 \int d^4x \sqrt{-g} \left( f(X) \, \Rf  + \sum_{A=1}^5 \alpha_A(X) \, L_A \right) \,,
\end{equation}
with
\begin{equation}
    \tilde{X} = \frac{X}{A+BX} \,,
\end{equation}
while the functions $f$ and $\alpha_A$ can be expressed in terms of $\tilde f$ and $\tilde\alpha_A$ as follows, 
\begin{equation}
    \begin{aligned}
        &f = \frac{J}{A} \tilde{f}, \quad \alpha_1 = -h + J T_{11} \tilde{\alpha}_1, \quad \alpha_2 = h + J T_{22} \tilde{\alpha}_2, \\
        & \alpha_3= 2 h_X + J (\tilde{f} \gamma_3 - 2 \tilde{f}_X \delta_3 + T_{13} \tilde{\alpha}_1 + T_{23} \tilde{\alpha}_2 + T_{33}\tilde{\alpha}_3) ,\\
        & \alpha_4= - 2 h_X + J (\tilde{f} \gamma_4 - 2 \tilde{f}_X \delta_4 + T_{14} \tilde{\alpha}_1 + T_{44} \tilde{\alpha}_4) ,\\
        & \alpha_5=  J (\tilde{f} \gamma_5 - 2 \tilde{f}_X \delta_5 + T_{15}\tilde{\alpha}_1 + T_{25} \tilde{\alpha}_2 + T_{35}\tilde{\alpha}_3 + T_{45} \tilde{\alpha}_4 + T_{55}\tilde{\alpha}_5) .
    \end{aligned}
\end{equation}
We have introduced the notations,
\begin{eqnarray*}
J = A^{3/2} \sqrt{A+BX}, \quad h = - \frac{B J \tilde{f}}{A(A+BX)},
\end{eqnarray*}
\begin{eqnarray*}
T_{11} & = &  \frac{1}{(A+B X)^2} \, , \quad T_{13} = \frac{2 A_X}{A (A+B X)^2} \, , \\
 T_{14} & = & \frac{2 \left(X \left(A_X+X B_X\right){}^2-A \left(2 \left(A_X+X B_X\right)+B\right)\right)}{A (A+B X)^3}, \\
 T_{15} & = & \frac{3 B^2 X^2 A_X^2}{A^2 (A+B X)^4}-\frac{2 B^2 X A_X}{A (A+B X)^4}+\frac{B^2}{(A+B X)^4}-\frac{2 B X^3 B_X^2}{A (A+B X)^4}-\frac{X^2 B_X^2}{(A+B X)^4} \\ 
&-& \frac{4 B X^2 A_X B_X}{A (A+B X)^4}+\frac{4 B X A_X^2}{A (A+B X)^4}-\frac{2 X A_X B_X}{(A+B X)^4}+\frac{4 B X B_X}{(A+B X)^4}+\frac{2 A_X^2}{(A+B X)^4}+\frac{2 A B_X}{(A+B X)^4} \, ,\\
T_{22} &= &\frac{1}{(A+B X)^2} \, , \quad T_{23} = -\frac{2 \left(A \left(-2 A_X+X B_X+B\right)-3 B X A_X\right)}{A (A+B X)^3}\, \\
T_{25} & = & \frac{\left(A \left(-2 A_X+X B_X+B\right)-3 B X A_X\right){}^2}{A^2 (A+B X)^4} \, ,\\
T_{33} & = & \frac{A-X \left(A_X+X B_X\right)}{(A+B X)^4} \, , \\
T_{35} & = & -\frac{\left(A \left(-2 A_X+X B_X+B\right)-3 B X A_X\right) \left(A-X \left(A_X+X B_X\right)\right)}{A (A+B X)^5}, \\
T_{44} & = & \frac{\left(A-X \left(A_X+X B_X\right)\right){}^2}{A (A+B X)^4} , \quad T_{45} = -\frac{B \left(A-X \left(A_X+X B_X\right)\right){}^2}{A (A+B X)^5}\, , \\
T_{55} &  = & \frac{\left(A-X \left(A_X+X B_X\right)\right){}^2}{(A+B X)^6} \, ,
\end{eqnarray*}
\begin{equation*}
    \begin{aligned}
    & \gamma_3 = -\frac{B \left(B X A_X+A \left(2 A_X+X B_X+B\right)\right)}{A^2 (A+B X)^2}, \\
    & \gamma_4 = \frac{A^2 \left(2 B A_X+4 X A_X B_X+6 A_X^2+B^2+B X B_X\right)+2 B^2 X^2 A_X^2+A B X A_X \left(8 A_X+4 X B_X+B\right)}{A^3 (A+B X)^2} ,\\
    & \gamma_5 = -\frac{2 A_X \left(B A_X+2 A B_X\right)}{A^3 (A+B X)} ,\\
    & \delta_3 = \frac{B}{A (A+B X)}, \quad \delta_4 = \frac{-4 B X A_X-A \left(6 A_X+2 X B_X+B\right)}{A^2 (A+B X)}, \quad \delta_5 = \frac{2 \left(2 B A_X+A B_X\right)}{A^2 (A+B X)} \,.
    \end{aligned}
\end{equation*}
All functions are evaluated in $X$ and not $\tilde{X}$ (and derivatives are with respect to $X$ and not $\tilde{X}$).

\bibliographystyle{utphys}
\bibliography{DHOST_geo}

\end{document}